\documentclass[aps,prl,groupedaddress,superscriptaddress,twocolumn]{revtex4}

\usepackage{amsmath}

\usepackage{graphicx}

\def\Vec#1{\mbox{\boldmath $#1$}}

\begin{document}
\title{Ferroelectricity induced by spin-dependent metal-ligand hybridization in  Ba$_2$CoGe$_2$O$_7$}

\author{H. Murakawa }
\affiliation{Multiferroics Project, ERATO, Japan Science and Technology Agency (JST), c/o Department of Applied Physics, University of Tokyo, Tokyo 113-8656, Japan}
 \altaffiliation{}
\author{Y. Onose}
\affiliation{Multiferroics Project, ERATO, Japan Science and Technology Agency (JST), c/o Department of Applied Physics, University of Tokyo, Tokyo 113-8656, Japan}
\affiliation{Department of Applied Physics, University of Tokyo, Tokyo 113-8656, Japan}

\author{S. Miyahara}
\affiliation{Multiferroics Project, ERATO, Japan Science and Technology Agency (JST), c/o Department of Applied Physics, University of Tokyo, Tokyo 113-8656, Japan}
 \altaffiliation{}
 
 \author{N. Furukawa }
\affiliation{Multiferroics Project, ERATO, Japan Science and Technology Agency (JST), c/o Department of Applied Physics, University of Tokyo, Tokyo 113-8656, Japan}
 \affiliation{Department of Physics and Mathematics, Aoyama Gakuin University, Sagamihara, Kanagawa 229-8558, Japan}

\author{Y. Tokura}

\affiliation{Multiferroics Project, ERATO, Japan Science and Technology Agency (JST), c/o Department of Applied Physics, University of Tokyo, Tokyo 113-8656, Japan}
\affiliation{Department of Applied Physics, University of Tokyo, Tokyo 113-8656, Japan}
\affiliation{Cross-Correlation Materials Research Group (CMRG) and Correlated Electron Research Group (CERG), RIKEN-ASI, Wako 351-0198, Japan}

\date{\today}

\begin{abstract}

We have investigated the variation of induced ferroelectric polarization under magnetic field with various directions and magnitudes in a staggered antiferromagnet Ba$_2$CoGe$_2$O$_7$.
While the ferroelectric polarization cannot be explained by the well-accepted spin current model nor exchange striction mechanism, we have shown that it is induced by the spin-dependent $p$-$d$ hybridization between the transition-metal (Co) and ligand (O) via the spin-orbit interaction. 
On the basis of the correspondence between the direction of electric polarization and the magnetic state, we have also demonstrated the electrical control of the magnetization direction.

\end{abstract}


\maketitle

Electrical control of magnetism has long been an important subject in condensed-matter physics as well as an urgent issue in contemporary spin-electronics. 
One of the promising ways toward this goal is to make use of magnetoelectric effect (change of magnetization by electric field, or reciprocally, change of electric polarization by magnetic field) in magnetic dielectrics\cite{Kimura,Fiebig,Cheong,Tokura,KimuraA}. 
The electric-field control of magnetism in terms of the magnetoelectric effect is much less dissipative in energy than the current control of magnetism in itinerant ferromagnets. 
While the magnetoelectric effect was thought to be quite small, the magnetically-induced ferroelectrics (multiferroics) and their giant magnetoelectric effect have recently been discovered and are now attracting much attention. 

There are several microscopic mechanisms for the magnetically-induced ferroelectricity. 
The most prevailing mechanism is the spin current mechanism\cite{Katsura}, or equivalently the inverse Dzyaloshinskii-Moriya mechanism\cite{Sel,Mos}. 
In the transverse-helical (cycloidal) spin structure, the spin chirality can induce the polarization $\Vec{P} \propto \sum_{i,j} \Vec{e}_{ij} \times (\Vec{S}_i \times \Vec{S}_i)$ in terms of the spin current mechanism, where $\Vec{e}_{ij}$ denotes the unit vector connecting the interacting neighbor spins $\Vec{S}_i$ and $\Vec{S}_j$. 
In the crystal which contains multiple inequivalent magnetic sites, even the collinear spin structure may also induce ferroelectricity with use of magnetostriction caused by the symmetric exchange interaction $J \Vec{S}_i \cdot \Vec{S}_j$\cite{ArimaL}. 
This type of the ferroelectricity is realized in perovskite $R$MnO$_3$ ($R$ = Y, Ho,..., Lu)\cite{Ishiwata}, DyFeO$_3$\cite{Tokunaga}, and Ca$_3$(Co,Mn)$_2$O$_6$\cite{Choi}. 
Quite recently, the third mechanism was proposed\cite{Arima} to explain the ferroelectricity induced by the proper screw or the 120$^{\circ}$ spin structure in delafossite CuFeO$_2$\cite{TKimura} and CuCrO$_2$\cite{KKimura}. 
In these materials, the ferroelectricity can be caused by the transition metal-ligand ($p$-$d$) hybridization depending on the spin direction\cite{Jia,Jia2}. 
Owing to the hybridization relevant to the spin-orbit interaction, the ionic charge $\rho$ of the ligand can vary depending on the angle $\eta$ between the spin of the transition metal and the vector $\Vec{e}$ connecting the transition metal and the ligand, i.e. $\Delta \rho \propto (\Vec{S} \cdot\Vec{e})^2$. 
Therefore, the local electric polarization $\Delta P \propto (\Vec{S} \cdot\Vec{e})^2\Vec{e}$ exists between the transition metal and the ligand. 
The induced charge of the local dipoles, when summed up over the whole crystal, may induce the ferroelectricity. 
To examine the mechanism in more detail, a more simple spin structure is desirable because this is a single ion problem rather than a spin correlation ($\Vec{S}_i \times \Vec{S}_j, \Vec{S}_i \cdot \Vec{S}_j$, etc.) one. 
In this paper, we show the ferroelectricity and its magnetic-field dependence in a simple staggered antiferromagnet Ba$_2$CoGe$_2$O$_7$ that can be throughly explained by such a spin-dependent $p$-$d$ hybridization mechanism. 
We also demonstrate a novel electrical response of magnetic properties based on this mechanism.

 \begin{figure}[t!]
\begin{center}
\includegraphics[width=0.9\linewidth]{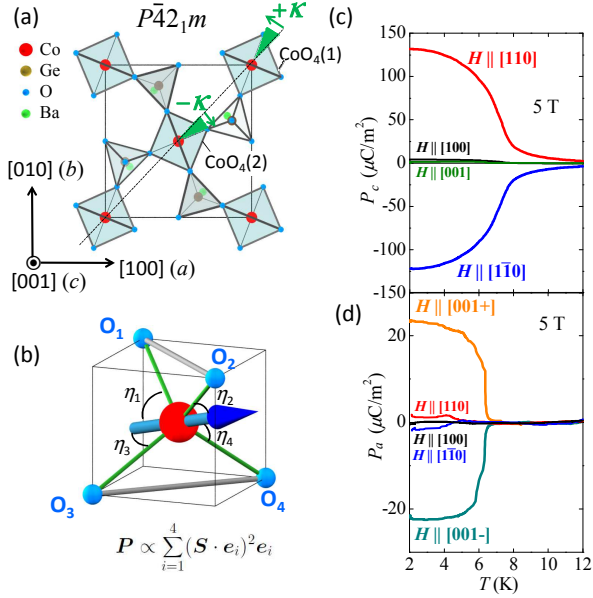} 
\caption[Structure]{(color online) (a) Schematic view of the crystal structure of Ba$_2$CoGe$_2$O$_7$.  
(b) Coordination of the spin moment in a CoO$_4$ tetrahedron.  
(c),(d) The temperature dependence of (c) [001] and (d) [100] components of the electric polarization $P$ in the magnetic field 5 T applied along various directions.  
  }
\label{1}
\end{center}
\end{figure}

Figure 1(a) illustrates the crystal structure of Ba$_2$CoGe$_2$O$_7$. 
It forms the tetragonal noncentrosymmetric structure (space group $P\bar{4}2_1m$), composed of corner shared CoO$_4$ and GeO$_4$ tetrahedra and intervening Ba$^{2+}$ ions. 
The Co magnetic moments have easy plane anisotropy and show a staggered antiferromagnetic structure in the (001) plane below $T_{\rm N}$ = 6.7 K\cite{Zheludev}. 
There are only two inequivalent Co sites and this is unchanged even below $T_{\rm N}$. 
Therefore, the sum of the vector spin chirality is zero and the $\Vec{S}_i \cdot \Vec{S}_j$ is uniform for all the Co-Co bond. 
For this reason, neither the $\Vec{S}_i \times \Vec{S}_j$ spin-current mechanism nor the $\Vec{S}_i \cdot \Vec{S}_j$ exchange striction one can work for this compound. 
However, the ferroelectric polarization is observed below $T_{\rm N}$ under magnetic field as reported for this\cite{Yi} and related compounds\cite{Akaki} (see also Figs. 1(c) and (d) observed in the present study), while the microscopic origin has not been clarified. 
In the following, therefore, we take the spin-dependent hybridization mechanism as the working hypothesis for the observed polarization.

According to the spin-dependent hybridization model, the local polarization of the CoO$_4$ tetrahedron can be expressed as $\Vec{P} \propto \sum_{i=1}^4(\Vec{S} \cdot \Vec{e}_i)^2\Vec{e}_i \propto \sum_{i=1}^4(S\cos\eta_i)^2 \Vec{e}_i$, where $\Vec{e}_i$ and $\eta_i$ are the vector connecting the Co and the $i$-th ligand O ions and the angle between the spin and $\Vec{e}_i$, respectively (Fig. 1(b)). 
When the spin moment $\Vec{S}$ is along the upper-lying oxygen bond O$_1$-O$_2$, the polarization appears along [001]. 
The polarization is unchanged after the 180$^{\circ}$ rotation of $\Vec{S}$, but is reversed when $\Vec{S}$ points along the lower-lying oxygen bond O$_3$-O$_4$. 
In Ba$_2$CoGe$_2$O$_7$, there are two inequivalent CoO$_4$ tetrahedra (CoO$_4$(1) and CoO$_4$(2)) of which the lower-lying oxygen bonds are tilted by the small angle $\pm \kappa$ from [110], respectively, as shown in Fig. 1(a). 
 When the Co spins in these two tetrahedra are respectively along ($\sin \theta_1 \cos \phi_1, \sin \theta_1 \sin \phi_1, \cos \theta_1$) and ($\sin \theta_2 \cos \phi_2, \sin \theta_2 \sin \phi_2, \cos \theta_2$) in the polar coordinate, the net polarization $\Vec{P} = (P_a,P_b,P_c)$ is expressed as  
\begin{gather} 
P_a \propto \sin 2\theta_1\sin (2\kappa - \phi_1) - \sin 2\theta_2 \sin (2\kappa + \phi_2), \notag \\
P_b \propto -\sin 2\theta_1 \cos (2\kappa - \phi_1) - \sin 2\theta_2\cos (2\kappa + \phi_2) , \\
P_c \propto \sin^2\theta_1\sin(2\kappa - 2 \phi_1) - \sin^2\theta_2\sin(2\kappa + 2\phi_2) . \notag
\end{gather}
\noindent The ferroelectric polarization and its magnetic-field dependence are well explained by this equation as discussed below. 

 \begin{figure}[b!]
\begin{center}
\includegraphics[width=0.9\linewidth]{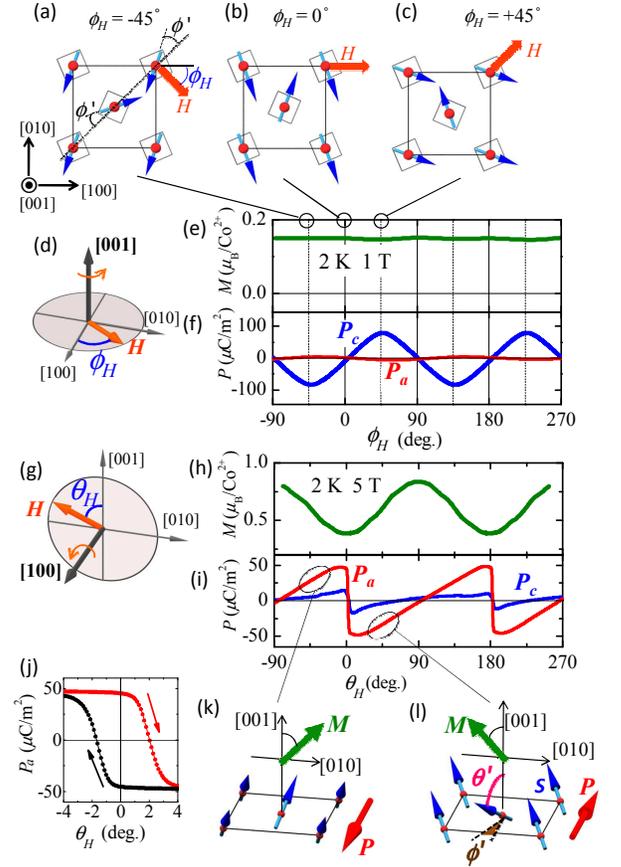} 
\caption{(color online) 
(a)-(c) Canted antiferromagnetic structures in magnetic fields with various in-plane directions. 
(d) The experimental configuration of the in-plane $H$-rotation measurement. 
(e),(f) The variations of (e) in-plane component of $M$ and (f) $P$ at 2 K under rotating $H$ around the [001] direction. 
(g) The experimental setup under rotating $H$ around the [100] direction. 
(h), (i) $\theta_H$ dependence of (h) $M$ and (i) $P$ at 2 K at 5 T. 
(j) The rotation hysteresis of $P_a$ at the vicinity of $H \parallel$ [001].
(k),(l) Canted antiferromagnetic structures and the induced electric polarization under out-of-plane magnetic field.   
 }
\label{2}
\end{center}
\end{figure}

We show the temperature dependence of $P$ along [001] ($P_c$) in Ba$_2$CoGe$_2$O$_7$ under various magnetic field ($H$) directions as shown in Fig. 1(c). 
In this measurement, $P$ was measured in a warming run after the $H$-cooling procedure but without poling procedure under electric field\cite{Poling}. 
When $H$ is applied along [110], $P$ emerges along the [001] direction. 
The sign of $P_c$ is reversed by changing the $H$ direction to [1$\bar{1}$0] and $P_c$ is almost zero under $H \parallel$ [100] and $H \parallel [001]$.  
Figures 2(e) and (f) show the variations of the magnetization ($M$) and $P$ at 2 K, respectively, with rotating $H$ in the (001) plane ($H = 1$ T). 
Here, $\phi_H$ is defined as the angle between [100] and the $H$ direction in the (001) plane as shown in Fig. 2(d). 
While the magnitude of $M$ is nearly invariant with $\phi_H$, $P_c$ shows sinusoidal $\phi_H$ dependence with the 180$^{\circ}$ period. 
The variation of $P$ with $\phi_H$ is in accord with the spin-dependent hybridization mechanism, as shown below.

In the magnetic field applied parallel to the magnetic easy (001) plane, the staggered magnetic moments tend to align nearly perpendicular to $H$ and canted to the direction of $H$ to induce $M$ as shown in Figs. 2(a)-(c). 
Judging from the nearly constant magnetic susceptibility with in-plane rotating $H$ in Fig. 2(e), there is almost no in-plane magnetic anisotropy and the staggered magnetic moments rotate freely in the (001) plane to keep $M$ parallel to $H$ as shown in Figs. 2(a)-(c). 
Therefore, in the case of the in-plane magnetic field, the spin directions $\theta_1$, $\theta_2$, $\phi_1$, $\phi_2$ can be expressed as $\theta_1 = \theta_2 = 90^{\circ}$, $\phi_1 = \phi_H - 90^{\circ} + \phi'$, $\phi_2 = \phi_H + 90^{\circ} - \phi'$, where $\phi'$ is the canting angle of the spin and invariant with $\phi_H$ (Fig. 2(a)). 
Putting these relations into Eq. (1), we obtain $\Vec{P} \propto (0,0,\sin2\phi_H \cos(2\kappa - 2\phi'))$. 
This relation clearly explains the sinusoidal $\phi_H$ dependence of the polarization with the 180$^{\circ}$ period shown in Fig. 2(f).

Next we proceed to the response of $P$ under out-of-plane magnetic field. 
The in-plane polarization in $H$ around [001] largely depends on the deviation of the $H$ direction from [001]. 
In Fig. 1(d), we show the temperature dependence of the polarization along [100] ($P_a$) in $H$ along various directions. 
When the magnetic field (5T) is applied along the direction slightly ($\sim 1^{\circ}$) slanted from [001] to [010] (denoted as $H \parallel [001+]$), the positive polarization appears below $T_{\rm N}$. 
When the slanted direction is reversed i.e. toward [0$\bar{1}$0] ($H \parallel [001-]$), the polarization is also reversed. 
The in-plane polarization is not observed under the in-plane magnetic field.  

Figures 2(h) and (i) show the variations of $M$ and $P$, respectively, at 2 K under rotating $H = 5$ T around [100]. 
The experimental set up is depicted in Fig. 2(g). 
Here $\theta_H$ is defined as the angle between the $H$ direction and the [001] direction. 
In contrast with the case of the in-plane rotation of $H$, $P$ is always along the [100] direction in this configuration (A small $P_c$ signal was caused perhaps by the misalignment of the magnetic-field direction).  
As shown in Fig. 2(h), $M$ shows sinusoidal $\theta_H$ dependence and takes a minimum value in $H \parallel [001]$, i.e. along the magnetic hard axis.  
On the other hand, $P_{a}$ shows the steep sign change around $\theta_H = 0^{\circ}$ and 180$^{\circ}$ and monotonically increases with the angle between the sign changes. 
Around $\theta_H = 0^{\circ}$ and 180$^{\circ}$, the rotational hysteresis is observed as shown in Fig. 2 (j). 

When the magnetic field is not parallel to the magnetic-easy $(001)$ plane ($\theta_H \neq 90^{\circ}$ or $270^{\circ}$), the canting angle of the staggered magnetic moments to the out-of plane direction is the same for the both sites ($\theta_1 = \theta_2 = \theta'$; see Fig. 2 (l)). 
The observed magnetic anisotropy in Fig. 2(h) indicates that the angle difference between $\theta'$ and $\theta_H$ $(0 \leq \theta_H \leq 180^{\circ})$ or $360^{\circ} - \theta_H$ $(180 < \theta_H \leq 360^{\circ})$ becomes larger as $\theta_H$ approaches 0$^{\circ}$ or 180$^{\circ}$ ($H \parallel [001]$). 
For the in-plane component of the staggered magnetic moments, the directions are expressed as $\phi_1 = \phi_H - 90^{\circ} + \phi', \phi_2 = \phi_H + 90^{\circ} - \phi'$, respectively.  
Here $\phi_H$ is the angle between the in-plane projected field and the [100] direction (Fig. 2(a)), and $\phi'$ is the in-plane canting angle (see Fig. 2(l) for definition). 
Putting these relations into Eq. (1), we obtain

$\Vec{P} \propto (\sin2\theta'\sin(2\kappa - \phi')\sin\phi_H, \sin2\theta'\sin(2\kappa-\phi')\cos\phi_H, \sin^2\theta'\cos(2\kappa - 2\phi')\sin2\phi_H)$. 

\noindent When the magnetic field is in the (100) plane, i.e. $\phi_H = 90^{\circ}$ ($180^{\circ} < \theta_H < 360^{\circ}$) or $270^{\circ}$ ($0^{\circ} < \theta_H < 180^{\circ}$), $\Vec{P} \propto (\sin2\theta'\sin(2\kappa-\phi'),0,0)$ for $\phi_H$ = 90$^{\circ}$ and $\Vec{P} \propto (-\sin2\theta'\sin(2\kappa-\phi'),0,0)$ for $\phi_H$ = 270$^{\circ}$. 
These relations clearly explain the sudden sign change of $P_a$ around $\theta_H = 0^{\circ}$ and 180$^{\circ}$. 
It is worth to note here that the observed $H$-direction dependence of polarization discussed so far is consistent with the symmetry analysis. 

 \begin{figure}[h!]
\begin{center}
\includegraphics[width=0.92\linewidth]{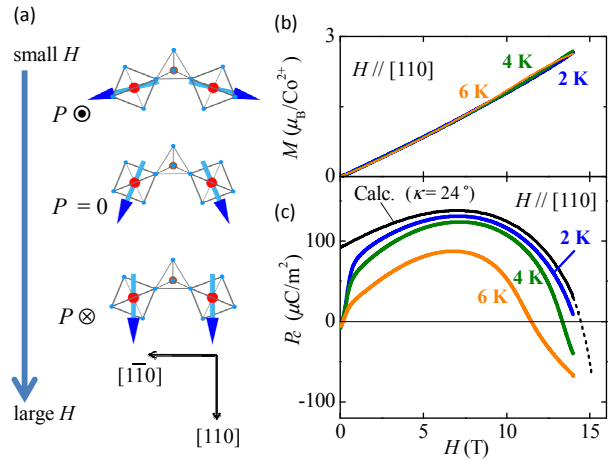} 
\caption{(color online)
(a)Illustration of the spin-structural change with increasing magnetic field along [110]. 
(b),(c) $H (\parallel [110])$ dependence of (b) $M$ (c) $P_c$ at various temperatures as well as the calculated curve of $P_c$ at 2 K based on the spin-dependent hybridization mechanism (see Eq. (1)). The definition of the structural parameter $\kappa$ is shown in Fig. 1(a). 
 }
\label{3}
\end{center}
\end{figure}

Now, we discuss the polarization as a function of $H$-magnitude to more decisively evidence the spin-dependent hybridization mechanism. 
Figure 3(c) shows the magnetic-field ($H \parallel [110]$) dependence of $P_c$ up to 14 T at various temperatures. 
The polarization increases with $H$ in the low-$H$ region and then decreases and shows the sign change ($\ge 4$ K) in the high-$H$ region. 
The critical magnetic fields of the downturn and sign change increase with decreasing temperature. 
The sign change is expected by the spin-dependent hybridization mechanism because the magnetic moments are almost perpendicular to $H$ in the low-$H$ region but parallel in the high-$H$ region, as shown in Fig. 3(a). 
In fact, the $c$-axis polarization can be expressed as $P_c \propto (1-2(M/M_{\rm S})^2)\cos 2\kappa + 2(M/M_{\rm S})(1-(M/M_{\rm S})^2)^{1/2}\sin 2\kappa$ according to the Eq. (1). 
Here, $M \propto \sin \phi'$ and $M_{\rm S}$ = 3 $\mu_{\rm B}$/Co$^{2+}$ is the saturation value of $M$.
As shown in Fig. 3(c), the observed magnetic-field dependence of $P_c$ at 2 K is well reproduced with the $M$ values measured at 2 K in Fig. 3(b) and the structural value $\kappa = 24^{\circ}$ of the isostructural compound Ca$_2$CoSi$_2$O$_7$\cite{Huaxue}, except for the low-$H$ region.  
In the low-$H$ region, the discrepancy between the observed and calculated values is large; the observed polarization is minimal at $H = 0$ while the calculated value is finite. 
Perhaps, the polarization spontaneously forms a multi-domain structure to minimize the electrostatic potential in the low-$H$ region.

 \begin{figure}[h!]
\begin{center}
\includegraphics[width=0.85\linewidth]{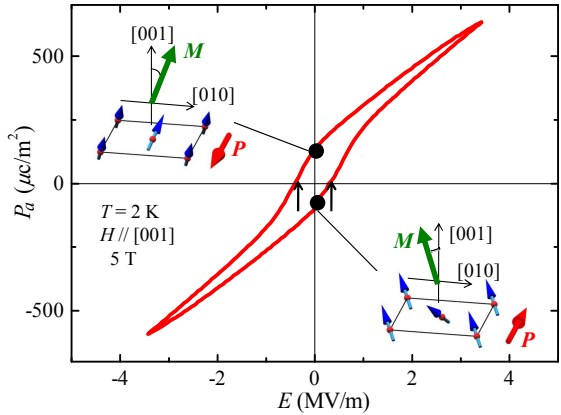} 
\caption{(color online) $P$-$E$ curve obtained in $H \parallel$ [001] of 5 T at 2 K. $P_a$ reversal is equivalent to the magnetization flop around [001] as shown in the insets. 
Incidentally, the large $E$-linear term in $P_a$ comes from the paraelectric component $\epsilon_{\rm r} \epsilon_0 E$, where $\epsilon_{\rm r} \sim 10$ and $\epsilon_0$ is the dielectric constant in the vacuum. }
\label{4}
\end{center}
\end{figure}

On the basis of the correspondence between the directions of $P$ and $M$ as clarified above, we demonstrate here a novel electric response of magnetism in Ba$_2$CoGe$_2$O$_7$. 
Figure 4 shows the $P$-$E$ curve obtained in $H \parallel [001]$ at 5 T at 2 K. 
Prior to the measurement, the $H$ direction was carefully aligned to [001]. 
In this case, two magnetic states shown in the inset of Fig. 4 are degenerate; the net magnetic moment is tilted from [001] to [010] directions in one state, while the tilting direction is reversed in the other state. (There are other degenerate states whose net magnetic moment is in the (010) plane. 
Nevertheless, these magnetic states cannot be controlled in terms of the electric field along [100].) 
We applied the electric field along [100] to obtain the clear $P$-$E$ curve.  
Since the sign of the transverse ($\parallel [010]$) component of the $M$ for $H \parallel [001]$ is tied to the in-plane $P$ direction as shown in Figs. 2 (k) and (l), the observed $P$ reversal indicates the magnetic domain switch as expressed by the inset scheme in Fig. 4.  
In this case, the $E$ reversal causes the large-$M$ ($\sim$ 0.4 $\mu_{\rm B }$/Co$^{2+}$) flop around the magnetic hard axis within 2$^{\circ}$.   
This is contrastive with the previous results of the electric-field control of magnetization for the canted antiferromagnet (Cu,Ni)B$_2$O$_4$ with a similar noncentrosymmetric crystal structure\cite{Saito}, where the electric field can switch the tilting angle of the much smaller magnetization ($ \sim 6 \times 10^{-3} \mu_{\rm B}/$Cu$^{2+}$) up to $\pm$30$^{\circ}$ from [100] within the magnetic-easy (001) plane.

In conclusion, we have studied the magnetoelectric properties of the staggered antiferromagnet with the noncentrosymmetric crystal structure Ba$_2$CoGe$_2$O$_7$. 
We have shown the $H$ dependent ferroelectric polarization can be explained in terms of the spin-dependent $p$-$d$ hybridization mechanism via the spin-orbit interaction. 
This is the first qualitative examination of this mechanism for multiferroicity. 
In addition, we have demonstrated the novel magnetoelectric functionalities such as the electric-field drive of the magnetization flop around the magnetic hard axis direction.

The authors thank I. Keszmarki, S. Seki and T. Arima for enlightening discussions. 
This work was in part supported by a Grant-In-Aid for Science Research from the Ministry of Education, Culture, Sports, Science and Technology (Grant nos. 20046004, 20340086,19684011,22014003), and by Funding Program for World-Leading
Innovative R\&D on Science and Technology (FIRST Program), Japan.


\begin{thebibliography}{99}
\bibitem{Poling}
This is because the degeneracy of the spin state can be lifted only by $H$ in this compound. This is contrastive with the case of ferroelectric helimagnets (except for Ba$_2$CuGe$_2$O$_7$\cite{Murakawa}), in which the spin helicity cannot be controlled by $H$ and the $E$-poling procedure is needed to observe $P$.
\end{thebibliography}
\end{document}